\documentclass[pra,aps,superscriptaddress,floatfix,tightenlines,twocolumn]{revtex4-1}

\usepackage{graphicx}
\usepackage{dcolumn}
\usepackage{bm}

\begin{document}

\title{Rayleigh scattering in an optical nanofiber as a probe of higher-order mode propagation}

\author{Jonathan E. Hoffman}
\affiliation{Joint Quantum Institute, Department of Physics, University of Maryland and National Institute of Standards and Technology, College Park, Maryland 20742, USA}
\author{Fredrik K. Fatemi$^*$}%
\affiliation{Optical Sciences Division, Naval Research Laboratory, Washington, DC, 20375}
\email{fredrik.fatemi@nrl.navy.mil}
\author{Guy Beadie}%
\affiliation{Optical Sciences Division, Naval Research Laboratory, Washington, DC, 20375}
\author{Steven L. Rolston}%
\affiliation{Joint Quantum Institute, Department of Physics, University of Maryland and National Institute of Standards and Technology, College Park, Maryland 20742, USA}
\author{Luis A. Orozco}%
\affiliation{Joint Quantum Institute, Department of Physics, University of Maryland and National Institute of Standards and Technology, College Park, Maryland 20742, USA}


\date{\today}



\begin{abstract}
Optical nanofibers provide a rich platform for exploring atomic and optical phenomena even when they support only a single spatial mode.  Nanofibers supporting higher-order modes provide additional degrees of freedom to enable complex evanescent field profiles for interaction with the surrounding medium, but local control of these profiles requires nondestructive evaluation of the propagating fields.  Here, we use Rayleigh scattering for rapid measurement of the propagation of light in few-mode optical nanofibers.  Imaging the Rayleigh scattered light provides direct visualization of the spatial evolution of propagating fields throughout the entire fiber, including the transition from core-cladding guidance to cladding-air guidance.  We resolve the interference between higher-order modes to determine local beat lengths and modal content along the fiber, and show that the modal superposition in the waist can be systematically controlled by adjusting the input superposition.  With this diagnostic we can measure variations in the radius of the fiber waist to below 3 nm \textit{in situ} using purely optical means.  This nondestructive technique also provides useful insight into light propagation in optical nanofibers.
\end{abstract}



\maketitle


\section{Introduction}

Recently, optical nanofibers (ONF) have received significant attention by realizing efficient interactions with atoms~\cite{Vetsch2010, OShea2013, Reitz2013, Morrissey2009, Nayak2007, LeKien2005, Pati, Morrissey2013, Mitsch2014, Lacroute2012, Polzik} and chemical analytes~\cite{Yalla2012, Fujiwara2011, GarciaFernandez}.  Knowledge of light propagation in ONFs is particularly important for atomic systems, for which the strong evanescent field enables both tightly trapped atoms near the nanofiber and efficient atom-light coupling for interrogation.  For a single-mode ONF that supports only the fundamental optical mode, the evanescent field distribution is well-defined and controlled; these ONFs have produced important results in atom-light interactions~\cite{Vetsch2010, OShea2013, Reitz2013, Morrissey2009, Nayak2007, LeKien2005}, but have been limited to only a few optical potential configurations.

To increase control over the evanescent field of the ONF and the complexity of the potential landscape, recent studies have explored the use of higher-order vector modes~\cite{Ravets2013a,Frawley2012,Sague2008b}, which have spatially-varying polarization profiles~\cite{Zhan2009,Fatemi2011}.   Although the ability to create, control, and detect complex optical mode distributions~\cite{Fatemi2011, Ravets2013a,Frawley2012, Fatemi2013, Golowich2013, Flamm2012}  at the input and output of a fiber is well-established, the challenge remains to analyze the propagation of light along the ONF taper and to measure the local optical intensities on the ONF waist.  In a single-mode ONF, the polarization dependence of Rayleigh scattering can discriminate between the two degenerate, quasi-linearly-polarized, fundamental modes along the fiber \cite{Vetsch2012, Marcuse1981, Goban2012,Lacroute2012, Eickhoff}, but such discrimination is not possible for higher-order modes (HOMs) without high transverse spatial resolution~\cite{Fatemi2015}.  Additionally, because ONFs are typically drawn to radii below the resolution limits of optical imaging, the ONF geometry is difficult to quantify \textit{in situ}.  Propagation dynamics through the fiber can be inferred from analysis of the transmitted light during the drawing process~\cite{Ravets2013a, Frawley2012}, but local visualization of the modes has not yet been achieved.

In this paper, we image Rayleigh scattered (RS) light to analyze light propagation in ONFs supporting the first six optical modes.  By combining high-spatial-resolution imaging with the polarization dependence of RS, we measure and control the propagation along the entire ONF length.  The intermodal beat lengths, which vary along these axially-tailored fibers, are quantified over the ONF and are sensitive to changes in the fiber radius to within 3 nm.  Finally, we show that by controlling the input modal superposition, we can systematically adjust the modal content on the ONF waist.

This paper is organized as follows. Section~\ref{subsec:RSMtheory} describes the dispersion relations in ONFs and properties of the modes of interest. Section~\ref{chap5:sec:ES_HOM} details the experimental setup used to fabricate and analyze propagation in the ONFs. In Sec.~\ref{sec:results} we present the key experimental results.  We present RS images over an entire ONF, including the  transition from core-cladding guidance to cladding-air guidance. We demonstrate modal identification and control of the mode distribution. We also discuss mode identification by further polarization discrimination of the RS light into longitudinal and transverse components.  Section 5 concludes the paper.

\bigskip

\section{Theory}\label{subsec:RSMtheory}

\begin{figure}[htb]
\centering
\includegraphics[width=8.4cm]{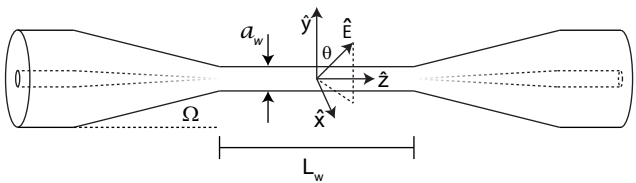}
\caption{Geometry of an optical nanofiber.  The camera viewing direction is along $\hat{y}$, and light propagates from left to right along $\hat{z}$.  $\hat{E}$ is the scattered field.  The tapered core is negligible in the fiber waist.
}
\label{fig:TaperGeometry}
\end{figure}

\subsection{Nanofiber geometry and modes}

An ONF consists of three sections: the unmodified fiber, the taper, and the fiber waist (See Fig.~\ref{fig:TaperGeometry}), with a symmetric taper on the output.  Initially we launch light into the core of the unmodified fiber on the left.  The taper makes an angle $\Omega$ with the fiber axis and adiabatically links the unmodified fiber to the fiber waist~\cite{Hoffman2014}.  Near the waist, the taper angle is reduced exponentially to connect the taper to the waist.  The waist has a sub-micrometer radius, $a_w$, and length, $L_{w}$ that is typically several millimeters long.

\begin{figure}[htb]
\centering
\includegraphics[width=8.4cm]{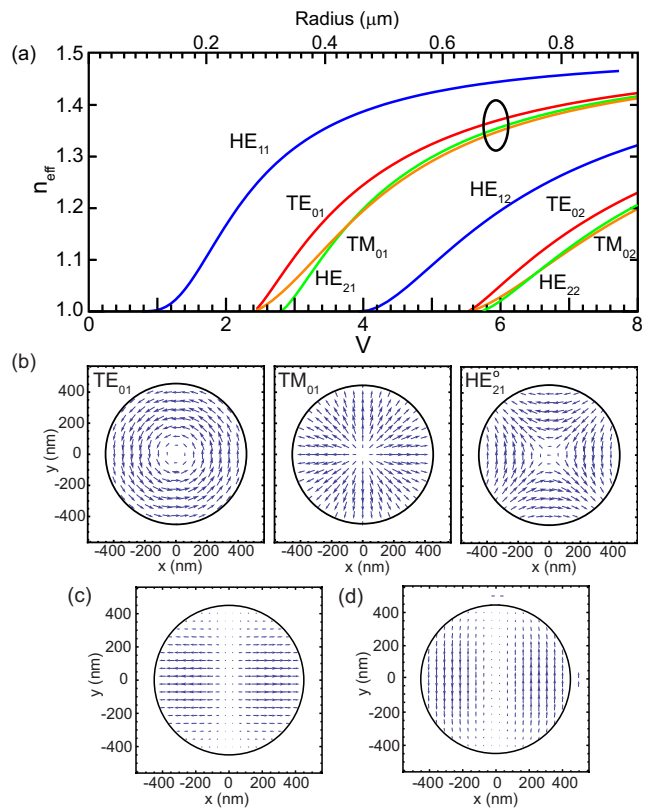}
\caption{ The effective index of refraction for the lowest lying modes in a nanofiber versus V-number (bottom axis) and radius (top axis). 
b) The vector profile inside a 450 nm radius fiber for the transverse components of the $TE_{01}$, $TM_{01}$ and $HE_{21}^o$ modes respectively.  The $HE_{21}^e$ mode is not shown, but is orthogonal at all points with the $HE_{21}^o$ mode. c) $HE_{21}^e$ and $TM_{01}$ modes interfering d) $HE_{21}^o$ and $TE_{01}$ modes interfering.  These images are representative of the transverse fields on the ONF waist.  The solid black circle marks the cladding air interface at a radius of 450 nm.  The refractive indices are chosen to be 1.45 and 1 for the cladding and air respectively. 
}
\label{fig:V}
\end{figure}

The number of guided fiber modes depends on the radius, $a$, the wavelength of input light, $\lambda$, and the refractive indices of the core and cladding, $n_1$ and $n_2$, through the dimensionless $V$-number:

\begin{equation}
V = a\, k \sqrt{n_1^2 - n_2 ^ 2},
\end{equation}

\noindent where $k=2\pi/\lambda$ is the free-space wave vector.  The allowed modes and their effective indices, $n_{\rm eff}$ are shown in Fig.~\ref{fig:V}(a) for an ONF with $V \leq 8$ ($a \leq 0.9~\mu$m).  In a tapered fiber, $V$ changes with propagation distance.  Previous works have examined taper geometries that allow adiabatic propagation to minimize interfamily mode-mixing~\cite{Ravets2013, Frawley2012}.  In this work, the ONFs are drawn from Fibercore SM1500 optical fiber, which has $V = 3.8$ for our operating wavelength $\lambda=$ 795 nm, and initially supports the $LP_{01}, LP_{11}$, and $LP_{21}$ mode families.  The $LP_{11}$ family is composed of the $TE_{01}$, $TM_{01}$, $HE_{21}^o$ and $HE_{21}^e$ vector modes, and their transverse field distributions are shown in Fig.~\ref{fig:V}(b).

Each mode has an associated propagation constant $\beta = n_{eff}k$.  Two propagating modes interfere with a beat length, $z_b$ defined by 

\begin{equation}\label{eqn:zb}
z_b(a) = \frac{2 \pi}{\beta_1(a) - \beta_2(a)} = \frac{\lambda}{n_{\rm eff,1}(a)-n_{\rm eff,2}(a)}.
\end{equation}

\noindent As we will show, the evolution of $z_b$ along the length of the fiber due to the tapered geometry is readily detected by imaging the RS light, and can be used to measure local fiber dimensions.

The propagation constants in Fig.~\ref{fig:V} have also been plotted as a function of $a$ (top axis).  From Eqn.~\ref{eqn:zb}, we can calculate the beat lengths for the modes of interest as a function of local fiber radius.  For an ONF with $a_w\approx$ 400 nm, $z_b$ varies from a few microns for HOMs beating with the fundamental mode, to tens of microns for HOMs beating together.  The $TE_{01}$ and $TM_{01}$ mode pair and the $HE_{21}^e$ and $HE_{21}^o$ mode pair do not interfere because their polarizations are orthogonal everywhere.

\bigskip

\subsection{Rayleigh scattering}

Standard optical fibers have birefringence due to strain, microbends, or deviations from cylindrical symmetry that can convert a pure input mode to a superposition of modes on the fiber waist.  This necessitates a nondestructive, \textit{in situ} measurement that locally probes the intensity and polarization.  Using Rayleigh scattering, we can extract information about the propagation and mode distribution along the ONF.  The observed RS power is proportional sin$^2\theta$, where $\theta$ is the angle between the local field, $\vec{E}$, and the camera viewing direction, $\hat{y}$.  When the mode-field-diameter is large enough, imaged RS shows the spatial structure of the propagating light~\cite{Fatemi2015}, and in Sec.~\ref{sec:results}\ref{sec:imaging} this will be used to observe the transition from core-cladding guidance to cladding-air guidance.  On the waist of the ONF, which typically has $a_w < 400$ nm, transverse mode structure cannot clearly be resolved.  However, since $z_b$ is much smaller than the length of the waist, the local superposition on the waist is determined straightforwardly through Eq.~\ref{eqn:zb}.
 
\bigskip
\section{Experimental setup}\label{chap5:sec:ES_HOM}

\subsection{Fiber pulling}

We use the flame brush technique to produce the ONFs~\cite{Birks1992, Warken2007b}. For the full details of the pulling apparatus see Refs.~\cite{Hoffman2014, Ravets2013,Ravets2013a}.  It consists of two computer-controlled high-precision motors and an oxyhydrogen flame in a stoichiometric mixture.  We employ an algorithm~\cite{Birks1992, Warken2007b, Ravets2013, Hoffman2014} to produce ONFs with linear tapers.  Following the protocols outlined in Ref.~\cite{Hoffman2014} we can produce  low-loss ONFs when controllably launching the fundamental mode~\cite{Ravets2013,Hoffman2014} or the next family of higher-order modes~\cite{Ravets2013a}.  Fiber cleanliness prior to tapering is crucial to achieving high transmission~\cite{Hoffman2014}.

\bigskip
\subsection{Mode preparation and imaging setup}

\begin{figure}[htb]
\centering
\includegraphics[width=8.4cm]{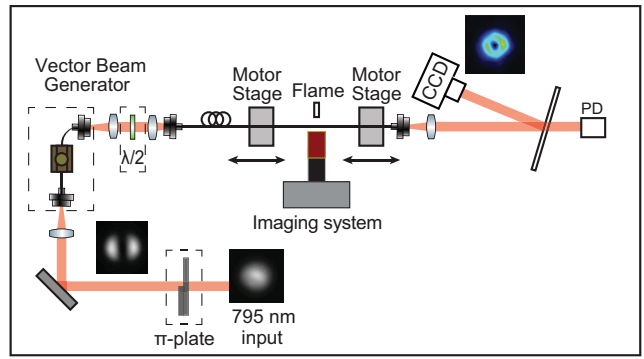}
\caption{ The experimental setup to measure the Rayleigh scattering from an ONF.}
\label{fig:HOM_expsetup}
\end{figure}

Figure~\ref{fig:HOM_expsetup} shows the experimental setup.  It has three components: the ONF fiber launch and vector beam generation setup~\cite{Fatemi2011, Ravets2013a}, the RS imaging setup, and the ONF output imaging setup.  The HOMs coupled into the ONF are derived from a fiber-based mode converter, described in detail in Ref.~\cite{Fatemi2011}.  Briefly, we begin with spatially-filtered, polarized light from a 795 nm diode laser.  This beam is collimated and passed through a phase plate that imparts a $\pi$ phase shift onto half of the beam, producing a two-lobed profile at the entrance of a vector beam generator.

The vector beam generator uses fiber that supports only the $HE_{11}$, $TE_{01}$, $TM_{01}$, and $HE_{21}$ modes at 795 nm (Corning HI1060).  A polarization controller (PC) mixes these modes through strain-induced birefringence and can produce either pure modes or any superposition of them at the output~\cite{Fatemi2013}.  Additional control over the mode superposition is done with bulk polarization optics (e.g. $\lambda$/2 waveplate in Fig.~\ref{fig:HOM_expsetup}) at the input of the ONF.  A CCD at the output of the ONF is used to aid alignment into the ONF.  The overall fiber length is 20-30 cm, and it is kept straight and tensioned slightly to reduce drifts.  We do not observe fluctuations over periods of several hours.

After we fabricate an ONF we use the motor stages (two high precision XML 210 motors see Fig.~\ref{fig:HOM_expsetup}) to translate the fiber relative to the RS imaging system to image the entire ONF.
RS light emitted orthogonal to the propagation direction is relayed to an EMCCD camera (Andor Luca) through a Mitutoyo M Plan APO NIR infinity-corrected objective (10x or 50x).  The imaging relay contains a 795 nm bandpass filter and a calcite walk-off polarizer near the CCD image plane to separate the longitudinal- and transverse-polarized RS light.  When using the calcite polarizer, the spatial resolution of the imaging system is limited to 2-3 $\mu$m.  Because the beat length between the $LP_{11}$ and $LP_{01}$ families is also 2-3 $\mu$m, this interference is not observed on the ONF waist.

\bigskip
\section{Results}
\label{sec:results}

\subsection{Direct imaging of optical propagation}\label{sec:imaging}

In a typical fiber, light is confined to the fiber core by the lower-index cladding, and the air surrounding the cladding has a negligible effect on the propagation.  On the ONF waist, however, the air itself becomes the cladding and the original fiber core is negligible.  The transition between these two regimes has previously only been indirectly studied by analyzing the transmitted power~\cite{Ravets2013a, Ravets2013, Frawley2012}.  In this section, we describe the direct imaging of the propagation evolution throughout the ONF.

\begin{figure}[htb]
\centering
\includegraphics[width=8.4cm]{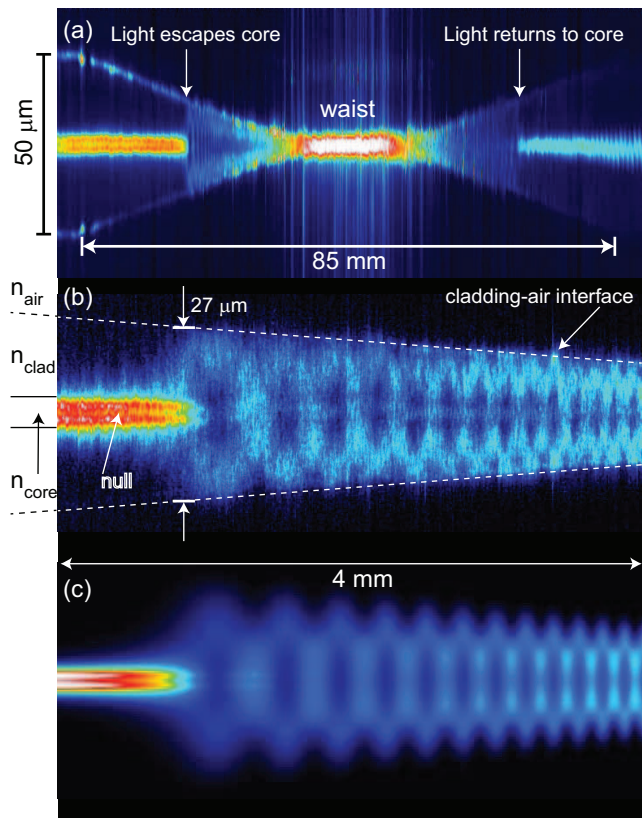}
\caption{High-resolution RS images of an ONF.  Light is propagating from left to right. (a) Propagation through the entire ONF, beginning and ending in unmodified fiber. $\Omega = 0.75$ mrad, $a_w = 1.5\mu$m.  b) Magnification of the input taper showing the transition from core to cladding near a radius of 13.5~$\mu$m.  c) Simulation of the propagation region shown in (b).  Colors in the images have been adjusted to highlight structure in the taper.}
\label{fig:prop_dyn}
\end{figure}

Using the experimental setup described in Sec.~\ref{chap5:sec:ES_HOM}, we launch a HOM into an ONF.  To observe the propagation over the entire 10-cm fiber length, we use the 10x objective, which has a field-of-view of $\approx$1 mm.  Figure~\ref{fig:prop_dyn}(a) shows a RS image produced after stitching together the series of images covering an ONF with $\Omega=$ 0.75 mrad and $a_w = 1.5~\mu$m.  In this figure, light is propagating from left to right.  The light that is initially in the core is ejected into the cladding when the fiber cladding has a diameter of 27 $\mu$m.  This is consistent with the value obtained indirectly in Ref.~\cite{Ravets2013a}.  On the output side, the field couples back into the core at the same radius, but the RS power is reduced due to light that escaped from the ONF on the input taper.  This light is lost because the ONF waist only supports the $LP_{01}$ and $LP_{11}$ mode families, whereas the unmodified fiber initially supported even higher order mode families at the fiber launch.  Additionally, higher-order mode families could be excited due to nonadiabatic propagation, as we have previously found that adiabaticity for the $LP_{11}$ family requires $\Omega < 0.4$ mrad~\cite{Ravets2013a}.   The color scaling in the image has been adjusted to highlight details in the taper and does not indicate saturation.  

Figure.~\ref{fig:prop_dyn}(b) is a magnification of Fig.~\ref{fig:prop_dyn}(a) over 4 mm of propagation distance in the critical region near $a=13.5~\mu$m, where the HOMs transition from the core to the cladding.  For these images we used a 50x objective (NA = 0.65) to collect the transverse polarization component of the RS light.  The core mode shows a null in the center of the scattering, which is an expected consequence of HOM Rayleigh scattering~\cite{Fatemi2015}, as superpositions of HOMs can have a two-lobed intensity pattern.  It is also possible that the field is a pure $HE_{21}^e$ or $TM_{01}$ mode, which would also appear with a null because of the polarization filtering of RS~\cite{Fatemi2015}.  After the core-cladding transition, the oscillations observed in Figs.~\ref{fig:prop_dyn}(a) and (b) increase in spatial frequency as the fiber tapers down to the ONF waist.  As will be discussed in Sec.~\ref{sec:results}\ref{chap5:sec:ss}, the total RS power increases toward the waist because of surface scattering effects.

We can analyze the length scales associated with the transition from core-cladding guidance to cladding-air guidance using the $1/e$ decay of the scattered power in the core.  This decay occurs over a propagation distance of $\approx$700 $\mu$m. For a fiber with $\Omega=0.75$ mrad this corresponds to a cladding radius change of 500~nm.  The specifications of SM1500 fiber indicate that the fiber has an initial core radius of 1.8 $\mu$m and cladding radius of 25 $\mu$m.  Tapering does not affect the core-to-cladding ratio, so this transition occurs over a change in core radius of $\approx$40 nm.  The oscillation pattern in the transition region is consistent with the length scales associated with the $LP_{11}$ family beating with the next symmetric family $LP_{21}$.  In this region, the relevant $V$-number governing the propagation increases rapidly from $V \approx 1.9$ before the transition to $V \approx 115$ after the transition.  Because the number of allowed modes scales with $V^2$, this is the region where intermodal transitions are most likely to occur and the adiabaticity criteria are most stringent~\cite{Ravets2013}.

Figure.~\ref{fig:prop_dyn}(c) is a simulation using commerical waveguide propagation software  (FIMMPROP/FIMMWAVE by Photon Design, Ltd.) of the local intensity, integrated along the viewing direction, over the region in Fig.~\ref{fig:prop_dyn}(b).  This calculated image is for a $TM_{01}$ mode, and does not include the polarization effects of RS, but the agreement between experiment and simulation is clear, including the location at which the light leaves the core and the chirped oscillation frequencies observed in the cladding.  In this region, the number of modes is large enough that it is possible to think in a pure ray optics picture, for insight, to illustrate the spatial oscillations of light and their change of frequency.

\bigskip
\subsection{Mode identification}

We can spatially resolve transverse modal information from RS when the light is confined to the core and when the ONF is larger than 3 $\mu$m, but direct imaging on the ONF waist is difficult, where $a_w \approx 400$ nm.  However, the oscillations observed in Fig.~\ref{fig:prop_dyn} are due to superpositions of modes that can also be identified by spatial frequency analysis of RS.  Such analysis is used in standard, single-mode birefringent fibers~\cite{Eickhoff}.  In the ONFs, however, several modes interfere with beat lengths that evolve along the fiber as the ONF radius, $a$, tapers.  We determine these $z_b$ through Fourier transform of the RS power as a function of $z$.  We use spectrogram analysis~\cite{Ravets2013,Orucevic2007} to obtain the inverse beat frequencies, $1/z_b$.  The spectrogram shows all the spatial frequencies that evolve as a function of propagation distance (and therefore radius) in the fiber.   

\begin{figure}[htb]
\centering
\includegraphics[width=8.4cm]{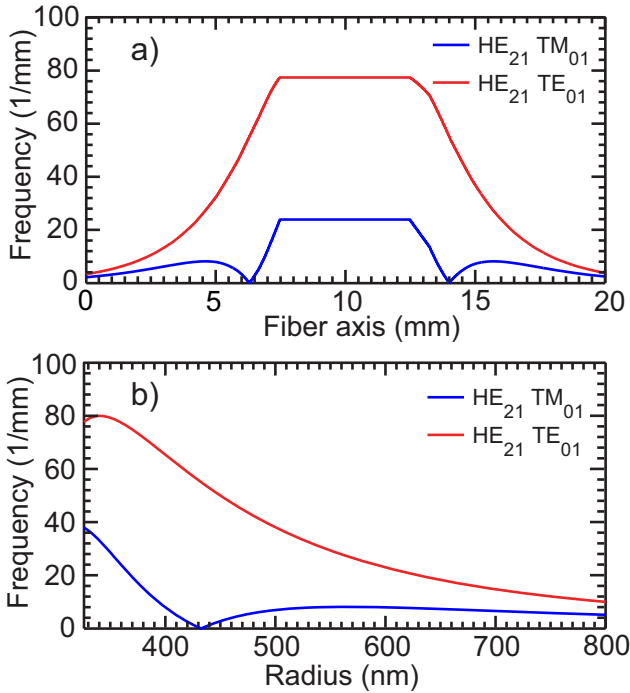}
\caption{Calculated inverse beat frequency for $a_w= 360$ nm with $\Omega = 1$ mrad and $L_w$ = 5 mm as a function of a) propagation distance; and b) ONF radius.}
\label{fig:sim_inversebeatlengths}
\end{figure}

Figure~\ref{fig:sim_inversebeatlengths}(a) shows the calculated inverse beat lengths as a function of position for a fiber with a profile $a_w = 360$ nm, $\Omega$ = 1 mrad, and $L_w$ = 5 mm using FIMMPROP. This is the shape we expect from the spectrogram for the $HE_{21}^o$ and $TE_{01}$ (red) beating together and the $HE_{21}^e$ and $TM_{01}$ (blue) interfering.  As shown below, the large quantitative and qualitative differences between these two curves enables us to identify the modes observed in the experiment.  The spatial frequencies for both superpositions have also been plotted as functions of $a$ in Fig.~\ref{fig:sim_inversebeatlengths}(b).

\begin{figure}[htb]
\centering
\includegraphics[width=8.4cm]{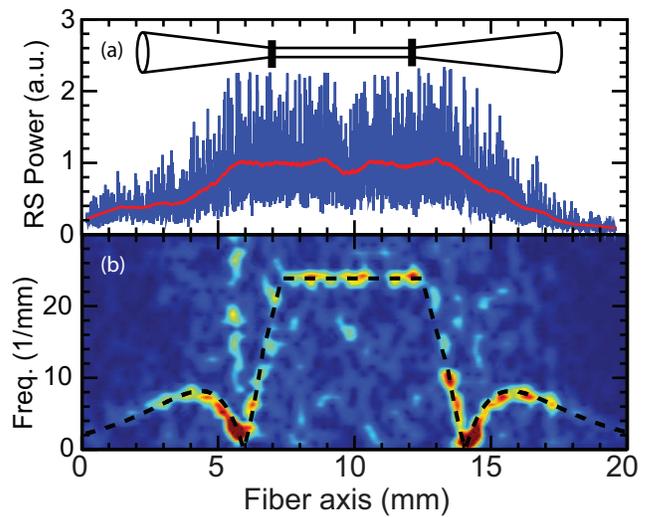}
\caption{ a) Detected (blue) and smoothed (red) transverse-polarized RS power along the ONF waist.  b) A spectrogram of concatenated images displaying the $HE_{21}$ mode beating with the $TM_{01}$ mode in arbitrary false color for the intensity.  The dashed black line corresponds to values expected from a fiber profile calculated by the pulling software using $a_w=360$ nm.}
\label{fig:HOM_scatpow}
\end{figure}
 
Fig.~\ref{fig:HOM_scatpow}(a) shows the power scattered as a function of $z$ near the ONF waist for a fiber with $\Omega=1$ mrad and a desired $a_w = 390$ nm.  The spatial variation in oscillation frequency of the RS power is depicted in the spectrogram in Fig.~\ref{fig:HOM_scatpow}(b), in which FFTs of the spatial domain data are calculated using 0.5 mm windows.  The qualitative shape in Fig.~\ref{fig:HOM_scatpow}(b) matches the shape of the $HE_{21}: TM_{01}$ curve from the calculation in Fig.~\ref{fig:sim_inversebeatlengths}a.  The spatial frequency is zero at the crossing of the $n_{\rm eff}$ curves for the $HE_{21}$ and $TM_{01}$ modes near 430 nm, shown in Fig.~\ref{fig:V}.  

The steep dependences of $1/z_b$ on $a$ are shown in Fig.~\ref{fig:sim_inversebeatlengths}(b) and allow a high-precision, local measure of variations in $a$.  We extend the FFT window over the entire waist to increase the spatial frequency resolution and obtain a beat frequency 1/$z_b$ = 24.50 $\pm$ 0.06~mm$^{-1}$ with a full-width at half-maximum (FWHM) of 0.24~mm$^{-1}$.  Considering the ONF as a single dielectric core surrounded by an infinite cladding of air, we estimate from this beat frequency that $a_w$ = 360 $\pm$ 10 nm.  With only the $HE_{21}^e:TM_{01}$ measurement, we cannot constrain $a_w$ with better than 10 nm uncertainty as it depends slightly on the exact index of the ONF.  However, the measured width of $1/z_b = 0.24$~mm$^{-1}$ limits the variation in $a_w$ to 0.7~nm over the entire 5 mm waist [see Fig.~\ref{fig:sim_inversebeatlengths}(b)].  Detailed calculations show that a similar high-resolution measurement of the beat frequency between $HE_{21}^o$ and $TE_{01}$ modes on the same fiber should constrain $a_w$, and this will be considered elsewhere.  The estimated value of $a_w=360$~nm is $\approx$30 nm smaller than the design radius of 390 nm.  Superimposed on the data in Fig.~\ref{fig:HOM_scatpow} is a beat frequency curve using a profile from the design software with a revised $a_w =$ 360  nm.  

\begin{figure}[htb]
\centering
\includegraphics[width=8.4cm]{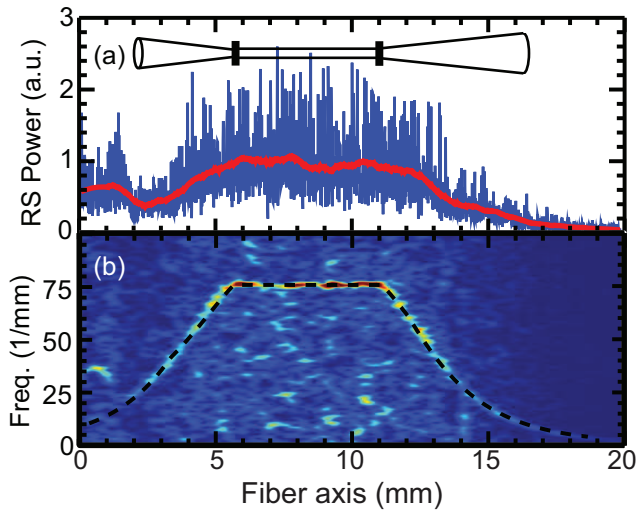}
\caption{ a) Detected (blue) and smoothed (red) transverse-polarized RS power along the ONF waist after summing over the transverse scattered power.  b) A spectrogram of aligned concatenated images displaying the $HE_{21}$ mode beating with the $TE_{01}$ mode in arbitrary false color for the intensity.  The dashed black line corresponds to an overlapped exponential fiber profile. The overlap results in a fiber radius of 370$ \pm 10$ nm.  Note this is for a different sample than the fiber displayed in Fig.~\ref{fig:HOM_scatpow}.}
\label{fig:HOM_scatpow_TE}
\end{figure}

Fig.~\ref{fig:HOM_scatpow_TE}(a) shows the RS profile for a different ONF sample than the data displayed in Fig.~\ref{fig:HOM_scatpow}. In this case, however, the shape of the spectrogram curve indicates beating between the $HE_{21}^o$ and $TE_{01}$ modes.  We follow a similar procedure for this ONF sample as the previous sample to extract a center beat frequency $1/z_b$ = 75.6 $\pm$ 0.1 mm$^{-1}$, with FWHM = 0.7~mm$^{-1}$.  Following a similar procedure as above, we estimate $a_w$ = 370 $\pm$ 10 nm.  At this radius, the beat frequency for the $HE_{21}^o:TE_{01}$ superposition has a weaker dependence on $a_w$ than the $HE_{21}^e:TM_{01}$ superposition and gives an estimated uniformity of 3 nm over the 5 mm length of the waist.  Beat frequency analysis provides a precise nondestructive extraction of the fiber profile without requiring high NA optics or irreversible coatings for imaging in a scanning electron microscope (SEM).

The total scattered power increases by 10x in Fig.~\ref{fig:HOM_scatpow} and 20x in Fig.~\ref{fig:HOM_scatpow_TE} as $a$ decreases, reaching a maximum on the ONF waist.  Because RS is directly proportional to intensity, the contribution from bulk RS should not increase.  As $a$ decreases, however, the field at the fiber surface increases, and it is likely that the observed increase in RS power comes from surface scattering.  This scattering appears as noise in the spectrograms and occurs at all spatial frequencies [see Fig.~\ref{fig:HOM_scatpow}(b) and Fig.~\ref{fig:HOM_scatpow_TE}(b)].  The surface scattering also increases at radii where HOMs reach cutoff.  This is apparent in Fig.~\ref{fig:HOM_scatpow_TE}(a), where the signal drops sharply near $z=2$ mm, at which point one of the $LP_{21}$ modes escapes the ONF.  On the output side of the fiber, this RS signature is absent because this HOM was no longer present.  Surface scattering will also be discussed in Sec.~\ref{sec:results}\ref{chap5:sec:ss}.

\bigskip
\subsection{Mode control} \label{Mode_Control}

The ability to measure spatial frequencies and identify participating modes suggests that external control of the input mode distribution can be used to create a desired mode superposition on the waist.  To demonstrate control we launch and manipulate the manifold of HOMs in a systematic way.  Following Ref.~\cite{Fatemi2011} we use the cylindrical vector beam generator and bulk optics, consisting of HWPs and polarizers, to launch known mode superpositions into the ONF and control the mode propagating on the ONF waist. 

Figure~\ref{fig:V}(b) displays the vector profile for the transverse components of a 450 nm radius fiber for the HOMs in the $LP_{11}$ family.  In free space, conversion between any of the pure modes is readily accomplished with one or two HWPs.  For example, by placing two HWPs with a relative angle of 45 degrees between their fast axes in front of a radially-polarized mode will produce an azimuthally-polarized mode at the fiber input.  Likewise, the $HE_{21}^e$ polarization profile is formed when a single HWP is placed in the path of a $TM_{01}$ profile.

Similarly, when launching an equal superposition of the $HE_{21}^e$ and $TM_{01}$ modes, which produces a two-lobed profile with polarization perpendicular to the null, we can completely convert to an equal superposition of $HE_{21}$ and $TE_{01}$ using a HWP with an angle of 45 degrees with respect to vertical.  Figs.~\ref{fig:V}(c) and \ref{fig:V}(d) show the transverse profiles of  the $HE_{21}^e:TM_{01}$ superposition and the $HE_{21}^o:TE_{01}$ superposition, respectively.  Beginning with the $HE_{21}^e:TM_{01}$ distribution in Fig.~\ref{fig:V}(c), it is straightforward to see that adding a HWP at angle $\alpha$ produces the $HE_{21}^o:TE_{01}$ superposition with relative power sin$^22{\alpha}$ and the $HE_{21}^e:TM_{01}$ superposition with relative power cos$^22\alpha$.  In a fiber, these superpositions may be modified due to strain or deviations from cylindrical symmetry.

\begin{figure}[htb]
\centering
\includegraphics[width=8.4cm]{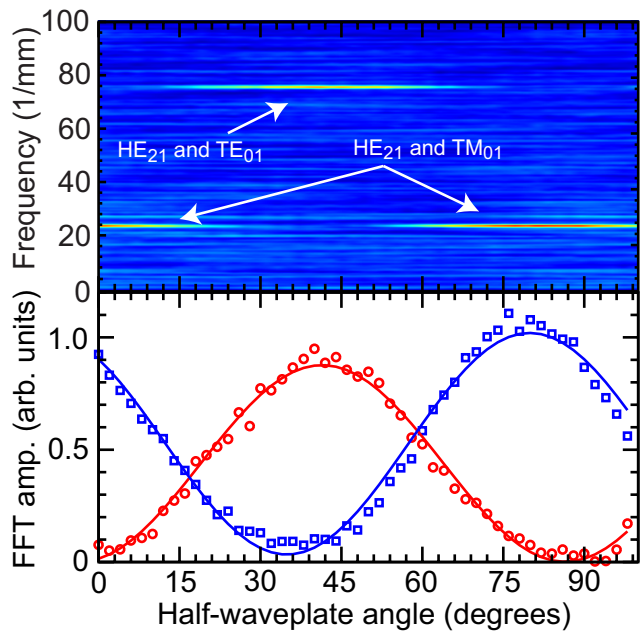}
\caption{Conversion from the $HE_{21}^{e}$ and $TM_{01}$ modes interfering to $HE_{21}^o$ and $TE_{01}$.  a) Each column corresponds to a FFT at a given HWP angle.  b) The absolute value of the FFT along the rows in (a) for the $HE_{21}^e$ and $TM_{01}$ beating together (blue squares) and the $HE_{21}^o$ and $TE_{01}$ modes beating together (red circles).  Fits to the expected sin$^22\alpha$ are shown as solid lines. 
}
\label{fig:mode_conversion_sum}
\end{figure}

Figure~\ref{fig:mode_conversion_sum} shows the relative powers of these superpositions as a function of $\alpha$.  Beginning with the $HE_{21}^o:TM_{01}$ input superposition, we rotate a HWP in 2 degree increments and collect the transverse RS over the ONF waist.  Then we take the absolute value of the Fourier transform at each $\alpha$ and plot it as the column of an image, as shown in in Fig.~\ref{fig:mode_conversion_sum}(a).  Note that the longitudinal ($\hat{z}$-polarized) RS cannot detect the $HE_{21}^o:TE_{01}$ superposition, because the $TE_{01}$ mode is entirely transverse.

The image shows two primary frequency bands, near $1/z_b$ = 75 mm$^{-1}$ and $1/z_b$ = 24~mm$^{-1}$, close to the two values shown in Fig.~\ref{fig:sim_inversebeatlengths} on the waist.  The powers of these individual frequency bands are also plotted in Fig.~\ref{fig:mode_conversion_sum}(b), with the $HE_{21}^e:TM_{01}$ shown in blue and the $HE_{21}^o:TE_{01}$ shown in red.  As described above, the data fit well to a sin$^22\alpha$ dependence, indicated by the solid lines.  At $\alpha = 0$, most of the signal is due to the $HE_{21}:TM_{01}$ superposition.  As $\alpha$ increases, the higher frequency band corresponding to the $HE_{21}^e:TM_{01}$ superposition increases and lower band corresponding to the $HE_{21}^o:TE_{01}$ superposition decreases.  At $\alpha\approx 40$ degrees, the $HE_{21}^o:TE_{01}$ superposition is maximized on the ONF waist.  Rotating the HWP another 45 degrees to 85 degrees minimizes the $HE_{21}^o:TE_{01}$ superposition.  The two bands are out of phase with almost full conversion from one pair to the other.  The small difference in amplitude of the two curves is due to the change in coupling into the ONF as the waveplate is rotated.

This control has been done entirely with bulk optics and a straight ONF.  For a fiber with bends, however, the transfer matrix between the input set of modes and the waist set of modes can have off-diagonal terms, requiring the ability to have arbitrary state generation at the input.  For vector modes, bulk optics alone cannot generate all possible input superpositions.  Polarization controllers relying on strain-induced birefringence, such as that used in the vector beam generator, can create any superposition of vector modes~\cite{Fatemi2013}. 

\bigskip
\subsection{Surface scattering and mode cutoffs} \label{chap5:sec:ss}

In Sec.~\ref{sec:results}\ref{Mode_Control}, we measured the modal superposition on the waist as a function of the input state by measuring spatial frequencies in the transverse-polarized RS data.  However, in these tightly confining waveguides, the propagating field can also have a longitudinal component of similar strength to the transverse component.  In this section, we analyze both the longitudinal and transverse components of the RS to enable more specific mode identification. This discrimination is done by adding a calcite walkoff plate near the camera plane to enable simultaneous measurement of both polarizations.  We also discuss the effects of surface scattering, which provides a useful, indirect measure of the field intensity at the nanofiber surface.

\begin{figure}[htb]
\centering
\includegraphics[width = 8.4cm]{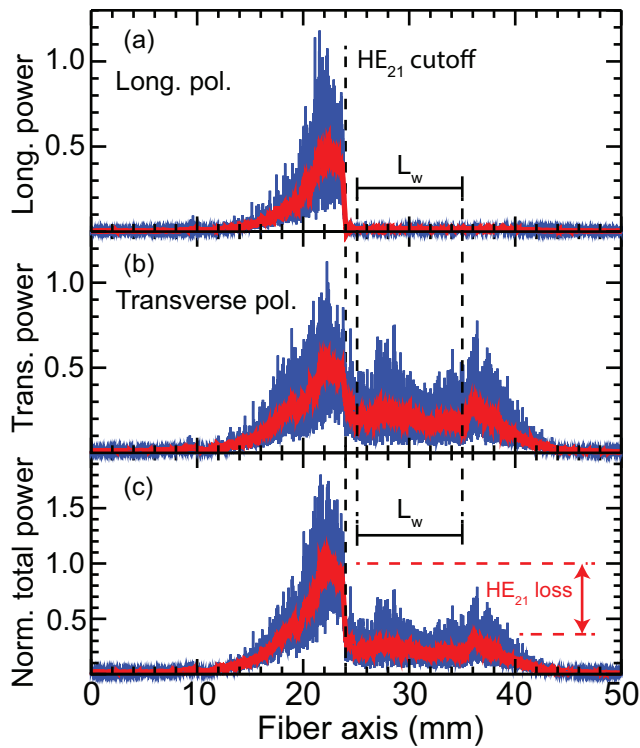}
\caption{Launching a superposition of HOMs into a fiber with $\Omega$ = 1 mrad, $a_w$ = 300 nm, and $L_{w}$ = 10 mm, marked with a thin, dashed black line.  The three panels correspond to scattering collected with (a) longitudinal polarization,  (b) transverse polarization, and (c) total scatter as a function of length respectively.  The long, thick dashed black line denotes the $HE_{21}$ cutoff and from that point we observe only the propagation of the $TE_{01}$ mode.  The short dashed red lines show the power loss from the $HE_{21}$ mode ejecting from the ONF.  The continuous black lines designate the ONF waist.  The power scattered from 0 mm to 25 mm corresponds to the input taper and the scattered power from 35 mm to 50 mm corresponds to the output taper. For clarity, the detected RS signal (solid blue) is smoothed (solid red) in each panel.}
\label{fig:cutoffs}
\end{figure}

As can be seen in Fig.~\ref{fig:V}, the $HE_{21}$ modes reach cutoff at a larger radius than the other modes in the $LP_{11}$ family.  By pulling a fiber to a radius just below the $HE_{21}$ cutoff, the only propagating modes are the $TE_{01}$ and $TM_{01}$ modes.  For these ONFs and 795 nm light, the $HE_{21}$ cutoff is reached at $a_w= 325$ nm.  Fig.~\ref{fig:cutoffs} shows the longitudinal, transverse, and total RS power from the ONF as a function of position along the fiber axis for $\Omega$ =  1 mrad, $a_w$ = 300 nm, and $L_w$= 1 cm.  By pulling to this radius, a number of phenomena are observed.

From $z\approx 0 - 10$ mm on the fiber axis, the total scattered power is constant.  That this region, a taper section, has a steady scattering rate, is indicative of bulk scattering.  After this point, from $z = 10$ mm to $z = 24$~mm, however, there is a steady increase in the scattering rate as the fiber continues to taper, consistent with surface scattering effects as discussed earlier.  The intensity of the mode at the fiber surface increases as the radius decreases and surface scattering begins to dominate over bulk scattering.  This increase in scattered power was also observed in Figs.~\ref{fig:HOM_scatpow}, and \ref{fig:HOM_scatpow_TE}.  

The RS light abruptly drops at $z\approx 25$ mm, indicated by the dashed line at $z=24$ mm in Fig.~\ref{fig:cutoffs}.  This is the point that the $HE_{21}$ mode reaches cutoff at $a = 325$ nm, corresponding to a $V\approx 2.8$.  The drop in RS power is most pronounced in the longitudinal polarization component in Fig.~\ref{fig:cutoffs}(a).   In each panel of Fig.~\ref{fig:cutoffs}, the RS light is constant between $z = 25$ mm and $z = 35$ mm.   This region, denoted $L_{w}$, is the 10-mm-long ONF waist.  The contribution of the $HE_{21}$ mode to the total scattered power is shown in Fig.~\ref{fig:cutoffs}(c), indicated by the red dashed lines.  These levels are taken at two symmetric places on opposite sides of the waist with equal radii.  

On the output side of the waist, the total RS power increases slightly.  This is again a consequence of surface scattering.  For this ONF, $a_w = 300$ nm and is also very close to the cutoff for the $TM_{01}$ and $TE_{01}$ modes.  At this radius, the remaining hollow modes have a significant evanescent field and low surface field.  As the radius increases on the output, the evanescent field reduces and the surface field increases, indicated by the larger RS power near $z=37$ mm.  The surface field reaches a maximum near $a=310$ nm, beyond which the field becomes more strongly confined inside the fiber and the surface scattering reduces. The RS power continues to drop as more of the field is confined to the bulk of the ONF.

The polarization discrimination of Fig.~\ref{fig:cutoffs}, combined with the cutoff of the $HE_{21}$ mode, helps identify the remaining amount of the $TM_{01}$ and $TE_{01}$ modes.  This is because the $TE_{01}$ mode has no longitudinal field and it should not contribute to any of the power seen in Fig.~\ref{fig:cutoffs}(a).  Since the $HE_{21}$ modes are ejected from the waveguide, and scattering still exists in the transverse direction but not in the longitudinal direction, we know that the majority of the total RS power on the waist is due to the $TE_{01}$ mode.  We note that there is some residual longitudinal component that could be caused by a small $TM_{01}$ or $HE_{11}$ background that is not observable in the spectrogram.  It could also be due to deviations from pure dipolar scattering~\cite{Mazumder}.  Ref.~\cite{Fatemi2015} notes that the longitudinal scattering was significantly larger than expected, even in unmodified fibers with negligible longitudinal fields.  These results indicate that pure $TE_{01}$ modes can be identified by minimizing the longitudinally-polarized RS light.

\bigskip
\section{Conclusion}\label{sec:conclusions}

We have shown that Rayleigh scattering is an excellent diagnostic of propagation phenomena in axially-tailored nanofibers.  Rayleigh scattering enables direct observation of the propagation throughout an optical nanofiber, allowing us to demonstrate control of the mode superpositions on the nanofiber waist.  By measuring the variation of the beat lengths, we can determine the local fiber radius to within 3 nanometers using entirely optical means.  The ability to measure and identify modes in high transmission, few-mode ONFs opens the possibility of controlling HOM-based evanescent traps for atoms.  This tool could also be extended to integrated waveguide approaches to atom traps~\cite{Lee2013}.

\bigskip
\section*{Funding Information}
This work was funded by Office of Naval Research, the ARO Atomtronics MURI, DARPA, and the NSF through the PFC at JQI.

\bigskip
\section*{Acknowledgments}

The authors gratefully acknowledge discussions with Sylvain Ravets.




\begin{thebibliography}{10}
\newcommand{\enquote}[1]{``#1''}

\bibitem{Vetsch2010}
E.~Vetsch, D.~Reitz, G.~Sagu\'e, R.~Schmidt, S.~T. Dawkins, and
  A.~Rauschenbeutel, \enquote{Optical interface created by laser-cooled atoms
  trapped in the evanescent field surrounding an optical nanofiber,} PRL
  \textbf{104}, 203603 (2010).

\bibitem{OShea2013}
D.~O'Shea, C.~Junge, J.~Volz, and A.~Rauschenbeutel, \enquote{Fiber-optical
  switch controlled by a single atom,} PRL \textbf{111}, 193601 (2013).

\bibitem{Reitz2013}
D.~Reitz, C.~Sayrin, R.~Mitsch, P.~Schneeweiss, and A.~Rauschenbeutel,
  \enquote{Coherence properties of nanofiber-trapped cesium atoms,} PRL
  \textbf{110}, 243603 (2013).

\bibitem{Morrissey2009}
M.~J. Morrissey, K.~Deasy, Y.~Wu, S.~Chakrabarti, and S.~Nic~Chormaic,
  \enquote{Tapered optical fibers as tools for probing magneto-optical trap
  characteristics,} Rev. Sci. Instrum. \textbf{80}, 053102 (2009).

\bibitem{Nayak2007}
K.~P. Nayak, P.~N. Melentiev, M.~Morinaga, F.~L. Kien, V.~I. Balykin, and
  K.~Hakuta, \enquote{Optical nanofiber as an efficient tool for manipulating
  and probing atomic fluorescence,} Opt. Express \textbf{15}, 5431--5438
  (2007).

\bibitem{LeKien2005}
F.~L. Kien, S.~D. Gupta, V.~I. Balykin, and K.~Hakuta, \enquote{Spontaneous
  emission of a cesium atom near a nanofiber: Efficient coupling of light to
  guided modes,} PRA \textbf{72}, 032509 (2005).

\bibitem{Pati}
S.~M. Spillane, G.~S. Pati, K.~Salit, M.~Hall, P.~Kumar, R.~G. Beausoleil, and
  M.~S. Shahriar, \enquote{Observation of nonlinear optical interactions of
  ultralow levels of light in a tapered optical nanofiber embedded in a hot
  rubidium vapor,} Phys. Rev. Lett. \textbf{100}, 233602 (2008).

\bibitem{Morrissey2013}
M.~J. Morrissey, K.~Deasy, M.~Frawley, R.~Kumar, E.~Prel, L.~Russell, V.~G.
  Truong, and S.~Nic~Chormaic, \enquote{Spectroscopy, manipulation and trapping
  of neutral atoms, molecules, and other particles using optical nanofibers: A
  review,} Sensors \textbf{13}, 10449 (2013).

\bibitem{Mitsch2014}
R.~Mitsch, C.~Sayrin, B.~Albrecht, P.~Schneeweiss, and A.~Rauschenbeutel,
  \enquote{Exploiting the local polarization of strongly confined light for
  sub-micrometer-resolution internal state preparation and manipulation of cold
  atoms,} PRA \textbf{89}, 063829 (2014).

\bibitem{Lacroute2012}
C.~Lacro\^ute, K.~S. Choi, A.~Goban, D.~J. Alton, D.~Ding, N.~P. Stern, and
  H.~J. Kimble, \enquote{A state-insensitive, compensated nanofiber trap,} New.
  J. Phys. \textbf{14}, 023056 (2012).

\bibitem{Polzik}
J.-B. B\'eguin, M.~Bookjans, E.\, L.~Christensen, S.\, L.~S\o{}rensen, H.\,
  H.~M\"uller, J.\, S.~Polzik, E.\, and J.~Appel, \enquote{Generation and
  detection of a sub-poissonian atom number distribution in a one-dimensional
  optical lattice,} Phys. Rev. Lett. \textbf{113}, 263603 (2014).

\bibitem{Yalla2012}
R.~Yalla, F.~L. Kien, M.~Morinaga, and K.~Hakuta, \enquote{Efficient channeling
  of fluorescence photons from single quantum dots into guided modes of optical
  nanofiber,} PRL \textbf{109}, 063602 (2012).

\bibitem{Fujiwara2011}
M.~Fujiwara, K.~Toubaru, T.~Noda, H.-Q. Zhao, and S.~Takeuchi, \enquote{Highly
  efficient coupling of photons from nanoemitters into single-mode optical
  fibers,} Nano Lett. \textbf{11}, 4362 (2011).

\bibitem{GarciaFernandez}
R.~Garcia-Fernandez, W.~Alt, F.~Bruse, C.~Dan, K.~Karapetyan, O.~Rehband,
  A.~Stiebeiner, U.~Wiedemann, D.~Meschede, and A.~Rauschenbeutel,
  \enquote{Optical nanofibers and spectroscopy,} Applied Physics B
  \textbf{105}, 3--15 (2011).

\bibitem{Ravets2013a}
S.~Ravets, J.~E. Hoffman, L.~A. Orozco, S.~L. Rolston, G.~Beadie, and F.~K.
  Fatemi, \enquote{A low-loss photonic silica nanofiber for higher-order
  modes,} Opt. Express \textbf{21}, 18325--18335 (2013).

\bibitem{Frawley2012}
M.~C. Frawley, A.~Petcu-Colan, V.~G. Truong, and S.~Nic~Chormaic,
  \enquote{Higher order mode propagation in an optical nanofiber,} Opt. Commun.
  \textbf{285}, 4648--4654 (2012).

\bibitem{Sague2008b}
G.~Sagu{\'e}, A.~Baade, and A.~Rauschenbeutel, \enquote{Blue-detuned evanescent
  field surface traps for neutral atoms based on mode interference in ultrathin
  optical fibres,} New J. Phys \textbf{10}, 113008 (2008).

\bibitem{Zhan2009}
Q.~Zhan, \enquote{Cylindrical vector beams: from mathematical concepts to
  applications,} Adv. Opt. Photon. \textbf{1}, 1--57 (2009).

\bibitem{Fatemi2011}
F.~K. Fatemi, \enquote{Cylindrical vector beams for rapid
  polarization-dependent measurements in atomic systems,} Opt. Express
  \textbf{19}, 25143--25150 (2011).

\bibitem{Fatemi2013}
F.~K. Fatemi and G.~Beadie, \enquote{Rapid complex mode decomposition of vector
  beams by common path interferometry,} Opt. Express \textbf{21}, 32291--32305
  (2013).

\bibitem{Golowich2013}
S.~Golowich, N.~Bozinovic, P.~Kristensen, and S.~Ramachandran, \enquote{Complex
  mode amplitude measurement for a six-mode optical fiber,} Opt. Express
  \textbf{21}, 4931--4944 (2013).

\bibitem{Flamm2012}
D.~Flamm, D.~Naidoo, C.~Schulze, A.~Forbes, and M.~Duparr\'{e}, \enquote{Mode
  analysis with a spatial light modulator as a correlation filter,} Opt. Lett.
  \textbf{37}, 2478--2480 (2012).

\bibitem{Vetsch2012}
E.~Vetsch, S.~T. Dawkins, R.~Mitsch, D.~Reitz, P.~Schneeweiss, and
  A.~Rauschenbeutel, \enquote{Nanofiber-based optical trapping of cold neutral
  atoms,} IEEE J. Sel. Top. Quant. Electron. \textbf{18}, 1763--1770 (2012).

\bibitem{Marcuse1981}
D.~Marcuse, \textit{Principles of optical fiber measurement} (Academic Press,
  1981).

\bibitem{Goban2012}
A.~Goban, K.~Choi, D.~J. Alton, D.~Ding, C.~Lacro\^{u}te, M.~Pototschnig,
  T.~Thiele, N.~P. Stern, and H.~J. Kimble, \enquote{Demonstration of a
  state-insensitive, compensated nanofiber trap,} Phys. Rev. Lett \textbf{109},
  033603 (2012).

\bibitem{Eickhoff}
W.~Eickhoff and O.~Krumpholz, \enquote{Determination of the ellipticity of
  monomode glass fibres from measurements of scattered light intensity,}
  Electron. Lett \textbf{12}, 405--407.

\bibitem{Fatemi2015}
F.~K. Fatemi and G.~Beadie, \enquote{Spatially-resolved rayleigh scattering for
  analysis of vector mode propagation in few-mode fibers,} Opt. Express
  \textbf{23}, 3831--3840 (2015).

\bibitem{Hoffman2014}
J.~E. Hoffman, S.~Ravets, J.~Grover, P.~Solano, P.~R. Kordell, J.~D.
  Wong-Campos, S.~L. Rolston, and L.~A. Orozco, \enquote{Ultrahigh transmission
  optical nanofibers,} AIP Advances \textbf{4}, 067124 (2014).

\bibitem{Ravets2013}
S.~Ravets, J.~E. Hoffman, P.~R. Kordell, J.~D. Wong-Campos, S.~L. Rolston, and
  L.~A. Orozco, \enquote{Intermodal energy transfer in a tapered optical fiber:
  optimizing transmission,} J. Opt. Soc. Am. A \textbf{30}, 2361--2371 (2013).

\bibitem{Birks1992}
T.~Birks and Y.~Li, \enquote{The shape of fiber tapers,} J. Lightwave Tech.
  \textbf{10}, 432--438 (1992).

\bibitem{Warken2007b}
F.~Warken, \enquote{Ultra thin glass fibers as a tool for coupling light and
  matter,} PhD Thesis,Rheinische Friedrich-Wilhelms Universitat, Mainz, Germany
  (2007) .

\bibitem{Orucevic2007}
F.~Orucevic, V.~Lef\`{e}vre-Seguin, and J.~Hare, \enquote{Transmittance and
  near-field characterization of sub-wavelength tapered optical fibers,} Opt.
  Express \textbf{15}, 13624--13629 (2007).

\bibitem{Mazumder}
P.~Mazumder, S.~L. Logunov, and S.~Raghavan, \enquote{Analysis of excess
  scattering in optical fibers,} Journal of Applied Physics \textbf{96},
  4042--4049 (2004).

\bibitem{Lee2013}
J.~Lee, D.~H. Park, S.~Mittal, M.~Dagenais, and S.~L. Rolston,
  \enquote{Integrated optical dipole trap for cold neutral atoms with an
  optical waveguide coupler,} New Journal of Physics \textbf{15}, 043010
  (2013).

\end{thebibliography}

\end{document}